\documentstyle[12pt]{article}
\title{ The dispersion relation of
        pion in nuclear matter }
\author{Liang-gang Liu  \\
CCAST (World Laboratory), P. O. Box 8730, Beijing 100080 \\
 {\scriptsize AND} \\
 Department of Physics, Zhongshan University \\
 Guangzhou, Guangdong 510275, P. R. China \\ 
Masahiro Nakano \\
University of Occupational and Environmental Health\\ 
  Kitakyusyu 807, Japan}
\date{}
\begin{document}
\maketitle
\vspace{2cm}

\begin{abstract}
We put forward a formalism to calculate the relativistic particle - hole and 
delta - hole excitation polarization insertion for pion propagator by using 
the particle - hole - antiparticle representation of nucleon and delta
propagators in nuclear matter. The real and the
imaginary part of the polarization insertion
and the dispersion relation for pion propagator are calculated numerically.
We find that the short range correlation enhances the delta - hole excitation
but suppresses the particle - hole excitation, it also suppresses the pion 
condensation. We find that the effect of the short range correlation on the
pion dispersion relation depends very much on the form of the short range
correlation and the parameters involved.
\end{abstract}

\newpage

\section{Introduction}
The pionic collective excitation and pion dispersion relation in nuclear
matter (or in finite nuclei) is very important to many fields in nuclear physics.
A typical example is the dilepton production in relativistic heavy - ion
collision in the energy region of $\rho$ meson production$^{1-4}$), 
where the production rate is determined by the pion 
dispersion relation and its derivative. The vanishing derivatives of the 
dispersion relation curve may lead to a huge enhancement of the dilepton
production rate.
In ref.$^{2}$), Xia et al calculated
the rate by using nonrelativistic pion dispersion relation which has very
good behavior for all momentum and energy. But it is not sure whether the 
nonrelativistic approximation is still applicable to very high energy - 
momentum transfer region. On the other hand, as far as pion dispersion 
relation is concerned, the imaginary part of the polarization insertion of 
pion propagator should not be ignored in the calculation.

Now, it is very clear that a microscopic study of the relativistic pion
dispersion relation is needed. The pion dispersion relation has been studied
extensively in the nonrelativistic nuclear matter$^{5-7}$), the nonrelativistic
particle - hole ({\it ph}) and delta - hole ($\Delta h$) excitation and short
range correlation via Landau - Migdal parameters $g'$ are taken into 
account. The main conclusion is that the pion condensation, or the space - like
branch of the pion dispersion relation in the critical region -- zero energy
and finite momentum, is removed by the short range correlation. 
Recently, there are some quasi-relativistic calculations on the $\Delta h$
excitation of the pion propagator at finite temperature$^{8, 9, 10}$).
In ref. 10, Helgesson and Randrup used a so called realistic $\pi$ + 
NN$^{-1}$ + $\Delta$N$^{-1}$ model to calculate the pion dispersion 
relation in 
nuclear matter, they found a very unusual but interesting result$^{10}$).
That is, besides the usual collective modes there is a number of modes they
named noncollective modes. One group of this noncollective modes shows pion
condensation. These are different from the previous results.
The real relativistic
calculation  can be found in a  recent comprehensive study$^{11}$). But as we
have pointed out and will show it later that by using "Feynman" and "Density 
dependent" nucleon propagator and Feynman propagator for delta isobars, the 
formulae for the relativistic {\it ph} and $\Delta h$ excitation can not be
derived correctly$^{12, 13}$). We have found a new method and solved this problem
$^{12, 13}$). This method has been used to study other properties of nuclear matter
, such as dimesonic function$^{14}$) and the binding energy of nuclear matter
$^{15, 16}$). Our method is as follows: to express nucleon propagator in nuclear
matter in terms of particle, hole, antiparticle propagators, and particle, 
antiparticle propagators for $\Delta$-isobars. Then the polarization insertion
can be explicitly split as {\it ph}, particle - antiparticle ($N \bar N$) and
$\Delta h$, antidelta - particle ($\bar\Delta N$), delta - antiparticle
($\Delta \bar N$) excitations contribution. The {\it ph} and $\Delta h$
excitations are finite, others are divergent and need a scheme to renormalize
it. The {\it ph} and $\Delta h$ are the most fundamental and important excitation
modes for pion propagator in nuclear matter, the thresholds for their physical
production are the lowest among all possible excitations. So it is 
necessary to
understand their contribution to the pion propagator first. In this paper, we
will study their polarization insertions, especially the imaginary part,
and pion dispersion relation.

In the next sect., we will derive the formulae for the relativistic {\it ph}
and $\Delta h$ excitation polarization insertion. In the 3rd sect., we give 
the numerical calculations especially on the  
polarization insertion and the dispersion relation of $\pi$ meson
with the emphasis on the effect of the short range correlation 
and their explanations. The summary
and conclusions will be given in the last section.

\section{Relativistic $ph$ and $\Delta h$ excitation}
In this section, we will give the real $ph$ and $\Delta h$ excitation 
polarization insertion for pion propagator.

\section*{2.1 Relativistic $ph$ excitation}
In ref.$^{12}$), we have shown how to derive the relativistic {\it ph} and
$N\bar N$ excitation polarization insertion in nuclear matter. In the following,
we will show the procedure briefly.

In nuclear matter, nucleon propagator $G(p)$ is usually written as "Feynman"
propagator $G_{F}(p)$ plus "Density dependent" part $G_{D}(p)$ $^{12}$). 
Taking pseudo-scalar (PS) $\pi NN$ coupling as an example: by using $G_{F}$ 
and $G_{D}$, the polarization insertion for pion propagator 
$\Pi^{PS} (\equiv \Pi^{PS}_{F} + \Pi^{PS}_{D})$ can be written as follows:
\begin{eqnarray}
\Pi^{PS}_{F}(q) & = &-  \frac{g^{2}_{\pi NN}}{(2\pi)^{3}} \int \frac{d {\bf p}}
{E_{{\bf p}}E_{{\bf p} - {\bf q}}} [4 E_{{\bf p} - {\bf q}} + \\ \nonumber
&  &q^{2} 
(\frac{1}{E_{{\bf p}} + E_{{\bf p} - {\bf q}} - q_{0} - i\epsilon} +
\frac{1}{E_{{\bf p}} + E_{{\bf p} - {\bf q}} + q_{0} - i\epsilon})],
\end{eqnarray}

\begin{eqnarray}
\Pi^{PS}_{D}(q)& =& \frac{g^{2}_{\pi NN}}{(2\pi)^{3}} \int \frac{d {\bf p}}
{E_{{\bf p}}E_{{\bf p} - {\bf q}}} \{(1 - n_{{\bf p}})n_{{\bf p} - {\bf q}}
\\ \nonumber &  & \cdot
q^{2}(\frac{1}{E_{{\bf p}} - E_{{\bf p} - {\bf q}} - q_{0} - i\epsilon} +
\frac{1}{E_{{\bf p}} - E_{{\bf p} - {\bf q}} + q_{0} - i\epsilon})  \\
\nonumber
&  & + n_{{\bf p}} [4 E_{{\bf p} - {\bf q}} + q^{2}(\frac{1}
.{E_{{\bf p}} + E_{{\bf p} - {\bf q}} - q_{0} - i\epsilon} +
\frac{1}{E_{{\bf p}} + E_{{\bf p} - {\bf q}} + q_{0} - i\epsilon}) ]\},
\end{eqnarray}
where $q = (q_0, {\bf q})$ is four energy-momentum,\footnote{We 
use the conventions of Bjorken and Drell, {\it Relativistic Quantum 
Fields}, (McGraw-Hill, New York, 1965).}
 $g_{\pi NN}$ is PS $\pi NN$ coupling constant. $E_{{\bf p}} = \sqrt{
{\bf p}^{2} + \tilde{m}^{2}_{N}}$, $\tilde{m}_{N}$ is nucleon effective mass 
in nuclear matter,
$n_{{\bf p}}$ is the nucleon distribution function at zero temperature. 
It is obvious $\Pi^{PS}_{F}$ is divergent, it comes from the contribution of
two $G_{F}$ in the integrand of the nucleon loop integration. The remaining 
part is finite and gives $\Pi^{PS}_{D}$. People usually omits $\Pi_{F}$ and
treat $\Pi_{D}$ as $ph$ excitation. Later we will see $\Pi^{PS}_{F}$ is not due to 
$N \bar N$excitation, neither $\Pi^{PS}_{D}$ is due to $ph$ excitation.

On the other hand, $G(p)$ can also be expressed as follows $^{17, 18}$):
\begin{eqnarray}
G(p) = S_{p}(p) + S_{h}(p) + S_{\bar{p}}(p)
\end{eqnarray}
\begin{eqnarray}
S_{p}(p) = \frac{\tilde{m}_{N}}{E_{{\bf p}}} \frac{(1 - n_{{\bf p}}) \Lambda_{+}(
{\bf p})}{p_{0} - E_{{\bf p}} + i \epsilon},
\end{eqnarray}
\begin{eqnarray}
S_{h}(p) =  \frac{\tilde{m}_{N}}{E_{{\bf p}}} \frac{ n_{{\bf p}} \Lambda_{+}(
{\bf p})}{p_{0} - E_{{\bf p}} - i \epsilon},
\end{eqnarray}
\begin{eqnarray}
S_{\tilde{p}}(p) = - \frac{\tilde{m}_{N}}{E_{{\bf p}}} \frac{ \Lambda_{-}(-
{\bf p})}{p_{0} + E_{{\bf p}} - i \epsilon},
\end{eqnarray}
here $\Lambda_{+}({\bf p})$, $\Lambda_{-}({\bf p})$ is the particle and 
antiparticle projection operator $S_{p}, S_{h}, S_{\bar{p}}$ is the particle, 
hole, antiparticle propagator, respectively. When these propagators are used to 
calculate polarization insertion, the nonvanishing parts will be  $ph$ and $N \bar N$
excitations, that is $\Pi^{PS}_{ph}, \Pi^{PS}_{N \bar N}$:
\begin{eqnarray}
\Pi^{PS} (q) = \Pi^{PS}_{ph} (q) +  \Pi^{PS}_{N \bar{N}} (q),
\end{eqnarray}
\begin{eqnarray}
 \Pi^{PS}_{ph} (q) &=& \frac{g^{2}_{\pi NN}}{(2\pi)^{3}}
\int \frac{d {\bf p}}{E_{{\bf p}}E_{{\bf p} - {\bf q}}} 
(1 - n_{{\bf p}})n_{{\bf p} - {\bf q}}[2(E_{{\bf p}} - E_{{\bf p} - {\bf q}})
\\ \nonumber
&  &+ q^{2}(\frac{1}{E_{{\bf p}} - E_{{\bf p} - {\bf q}} - q_{0} - i\epsilon} +
\frac{1}{E_{{\bf p}} - E_{{\bf p} - {\bf q}} + q_{0} - i\epsilon})],
\end{eqnarray}
\begin{eqnarray}
 \Pi^{PS}_{N \bar{N}} (q) & = &- \frac{g^{2}_{\pi NN}}{(2\pi)^{3}}
\int \frac{d {\bf p}}{E_{{\bf p}}E_{{\bf p} - {\bf q}}} 
(1 - n_{{\bf p}})[2(E_{{\bf p}} + E_{{\bf p} - {\bf q}})
\\ \nonumber
&  &+ q^{2}(\frac{1}{E_{{\bf p}} + E_{{\bf p} - {\bf q}} - q_{0} - i\epsilon} +
\frac{1}{E_{{\bf p}} + E_{{\bf p} - {\bf q}} + q_{0} - i\epsilon})],
\end{eqnarray}

The sum of $\Pi^{PS}_{ph}$ and $\Pi^{PS}_{N \bar N}$ is equal to sum of
$\Pi^{PS}_{F}$ and $\Pi^{PS}_{D}$. Since $\Pi^{PS}_{ph}$ is not equal to 
$\Pi^{PS}_{D}$, $\Pi^{PS}_{N \bar N}$ is not equal to $\Pi^{PS}_{F}$, so 
$\Pi^{PS}_{F}$ does not stand for $N \bar N$ excitation, neither $\Pi^{PS}_{D}$
stands for $ph$ excitation. That means one can not get the correct expressions for 
$ph$ and $N \bar N$ excitations if $G_{F}, G_{D}$ propagators are used to 
calculate the polarization insertion.

Similarly, in case of pseudo-vector (PV) $\pi NN$ coupling, the $ph$ and $N \bar N$
excitation polarization insertions $\Pi^{PV}_{ph}, \Pi^{PV}_{N \bar N}$ can be
written as follows:
\begin{eqnarray}
\Pi^{PV}_{ph} (q) & = &\frac{2f^{2}_{\pi NN}}{(2\pi)^{3}m^{2}_{\pi}}
\int \frac{d {\bf p}}{E_{{\bf p}}E_{{\bf p} - {\bf q}}} 
(1 - n_{{\bf p}}) n_{\bf p - q} \{(E_{{\bf p} - {\bf q}} - E_{\bf p}) 
[(E_{{\bf p}} + E_{{\bf p} - {\bf q}})^{2} - {\bf q}^{2}] 
\\ \nonumber
&  &+ 2 {\tilde m}^{2}_{N} q^{2}(\frac{1}{E_{{\bf p}} - 
E_{{\bf p} - {\bf q}} - q_{0} - i\epsilon} +
\frac{1}{E_{{\bf p}} - E_{{\bf p} - {\bf q}} + q_{0} - i\epsilon})\},
\end{eqnarray}
\begin{eqnarray}
\Pi^{PV}_{N \bar{N}} (q) & = &\frac{2f^{2}_{\pi NN}}{(2\pi)^{3}m^{2}_{\pi}}
\int \frac{d {\bf p}}{E_{{\bf p}}E_{{\bf p} - {\bf q}}} 
(1 - n_{{\bf p}}) \{(E_{{\bf p}} + E_{{\bf p} - {\bf q}}) 
[(E_{{\bf p}} - E_{{\bf p} - {\bf q}})^{2} - {\bf q}^{2}]
\\ \nonumber
&  &- 2 {\tilde m}^{2}_{N}q^{2}(\frac{1}{E_{{\bf p}} + 
E_{{\bf p} - {\bf q}} - q_{0} - i\epsilon} +
\frac{1}{E_{{\bf p}} + E_{{\bf p} - {\bf q}} + q_{0} - i\epsilon})\},
\end{eqnarray}
here $f_{\pi NN}$ = 0.988 is PV $\pi NN$ coupling constant$^{11}$). The 
analytic expression for the real and the imaginary part of eq. (10) can be
found in ref.$^{12}$).

\section*{2.2 Relativistic $\Delta h$ excitation}
The delta baryon Feynman propagator $S^{\mu \nu}(p)$ can be expressed in 
terms of particle and antiparticle propagators, that is $S^{\mu \nu}_{\Delta}
(p)$ and $S^{\mu \nu}_{\bar \Delta}(p)$:
\begin{eqnarray}
S^{\mu \nu}(p) = S^{\mu \nu}_{\Delta}(p) + S^{\mu \nu}_{\bar \Delta}(p),
\end{eqnarray}
\begin{eqnarray}
S^{\mu \nu}_{\Delta}(p) = \frac{{\tilde m}_{\Delta}}{E_{\Delta}({\bf p})}
\cdot \frac{\Lambda^{\mu \nu}_{+}({\bf p})}{p_{0} - E_{\Delta}({\bf p}) 
+ i\epsilon},
\end{eqnarray}
\begin{eqnarray}
S^{\mu \nu}_{\bar \Delta}(p) = - \frac{{\tilde m}_{\Delta}}{E_{\Delta}({\bf p})}
\cdot \frac{\Lambda^{\mu \nu}_{-}( - {\bf p})}{p_{0} + E_{\Delta}({\bf p}) 
- i\epsilon},
\end{eqnarray}
here $E_{\Delta}({\bf p}) = \sqrt{{\bf p}^{2} + {\tilde m}^{2}_{\Delta}}$, 
${\tilde m}_{\Delta}$ is effective mass of $\Delta$-isobars in nuclear matter.
(Here we implicitly assume there is no real $\Delta$-isobars in the ground
state, otherwise the decay of a real $\Delta$-isobar will cause the instability of nuclear 
matter$^{19}$), for this reason we can omit the width of $\Delta$-isobars.)
$\Lambda^{\mu \nu}_{+}({\bf p})$, $\Lambda^{\mu \nu}_{-}({\bf p})$ is particle
and antiparticle projection operator, respectively,
\begin{eqnarray}
\Lambda^{\mu \nu}_{+}({\bf p}) = - \frac{\gamma \cdot p + {\tilde m}_{\Delta}}
{2 {\tilde m}_{\Delta}} \cdot [g^{\mu \nu} - \frac{1}{3} \gamma^{\mu} 
\gamma^{\nu} - \frac{2}{3 {\tilde m}^{2}_{\Delta}}p^{\mu}p^{\nu} + \frac{1}
{3 {\tilde m}_{\Delta}}(p^{\mu}\gamma^{\nu} - p^{\nu}\gamma^{\mu})],
\end{eqnarray}
\begin{eqnarray}
\Lambda^{\mu \nu}_{-}({\bf p}) = - \frac{- \gamma \cdot p + {\tilde m}_{\Delta}}
{2 {\tilde m}_{\Delta}} \cdot [g^{\mu \nu} - \frac{1}{3} \gamma^{\mu} 
\gamma^{\nu} - \frac{2}{3 {\tilde m}^{2}_{\Delta}}p^{\mu}p^{\nu} - \frac{1}
{3 {\tilde m}_{\Delta}}(p^{\mu}\gamma^{\nu} - p^{\nu}\gamma^{\mu})],
\end{eqnarray}
here $p = (E_\Delta({\bf p}), {\bf p})$. There is a discrepancy on the overall
sign for the free Lagrangian or the propagator of $\Delta$-isobar$^{20}$).
We found and checked that an overall negative sign is very important to keep
the imaginary part of $\Delta h$ excitation polarization insertion negative.

The $\pi N \Delta$ interaction is given as follows$^{11}$):
\begin{eqnarray}
{\cal L}_{\pi N \Delta} = \frac{f_{\pi N \Delta}}{\sqrt{2} m_{\pi}}
{\bar \psi}^{\mu}_{\Delta}(g_{\mu \nu} + \xi \gamma_{\mu} \gamma_{\nu})
{\vec T} \cdot \psi_{N} \partial^{\nu}{\vec \pi} + h.c.,
\end{eqnarray}
here $f_{\pi N \Delta}$ = 2$f_{\pi NN}$, $\xi$ is an arbitrary constant
$^{20, 21}$). It will contribute to the processes in which $\Delta$-isobar
is off-shell.
Since $\gamma_{\mu} \Lambda^{\mu \nu}_{+}({\bf p})$ = $\Lambda^{\mu \nu}_{+}
({\bf p}) \gamma_{\nu}$ = 0, so $\xi$ will not appear in the polarization function.

The nucleon-delta excitation is a very important channel to $\pi$ propagator,
it is nothing but $\Delta h$, $\Delta \bar N$, 
${\bar \Delta} N$ excitation. In this paper we only consider $\Delta h$ 
excitation. The excitation of other two channels require higher energy.
Using $S_{h}$, $S^{\mu \nu}_{\Delta}$, ${\cal L} _{\pi N \Delta}$ 
and Feynman rules, $\Delta h$ polarization insertion can be given as follows:
\begin{eqnarray}
{\hskip 0.1cm}\Pi_{\Delta h}(q_{0}, {\bf q}) & = & - 
\frac{4f^{2}_{\pi N \Delta}}{9 {\tilde m}^{2}_{\Delta}
m^{2}_{\pi}} \int \frac{d {\bf p}}{E_{\Delta}({\bf p})E_{{\bf p - q}}} 
n_{\bf p - q} \{ 2(E_{\Delta}({\bf p}) - E_{\bf p - q}) 
[E_{\Delta}({\bf p})E_{\bf p - q} 
q^{2}_{0} 
\\ \nonumber
&  &+ {\tilde m}^{2}_{\Delta} q^{2} - ({\bf p} \cdot {\bf q})^{2}]
- 2 ({\tilde m}^{2}_{\Delta} - {\tilde m}^{2}_{N} - {\bf q}^{2})
E_{\Delta}({\bf p}) q^{2}_{0} 
+[({\tilde m}_{\Delta} + {\tilde m}_{N})^{2}- q^{2}] \cdot
\\ \nonumber
&  &[\frac{1}{2} (
(E_{\Delta}({\bf p}) + E_{\bf p - q} )^{2} + q^{2}_{0})
(E_{\Delta}({\bf p}) - E_{\bf p - q} ) + ({\tilde m}^{2}_{N} + {\bf q}^{2} -
{\tilde m}^{2}_{\Delta})(E_{\Delta}({\bf p}) + E_{\bf p - q} ) ]
\\ \nonumber
&  & + [q^{2} - ({\tilde m}_{\Delta} + {\tilde m}_{N})^{2}]
[{\tilde m}^{2}_{\Delta} q^{2} - \frac{1}{4}
({\tilde m}^{2}_{\Delta} - {\tilde m}^{2}_{N} +  q^{2})^{2}] \cdot
\\ \nonumber
&  & (\frac{1}{E_{\Delta}({\bf p}) - E_{\bf p - q} 
- q_{0} - i\epsilon} +
\frac{1}{E_{\Delta}({\bf p}) - E_{\bf p - q} + q_{0} - i\epsilon}) \}.
\end{eqnarray}

As we have mentioned above, $\xi$ does not appear. But this is not the case
if we use $G_{D}$ and Feynman propagator $S^{\mu \nu}$$^{11}$).

The short range correlation  between $NN, N\Delta, \Delta\Delta$ can be
included via Landau-Migdal parameters     
$g'_{NN}$, $g'_{N \Delta}$, $g'_{\Delta \Delta}$. 
But the way of constructing short range correlation interacting vertex and 
intermediate state propagator is not unique$^{12}$). Here we will 
use the method as other papers used$^{2, 11}$). 
Taking into account this effect, the polarization 
insertion $\Pi '$ which includes $ph$, $\Delta h$ excitation and short range correlation in the Random Phase
Approximation, can be written as follows$^{11}$):
\begin{eqnarray}
\Pi'(q) = q^{2} \frac{q^2(\Pi^{PV}_{ph}(q) + \Pi_{\Delta h}(q)) + 
\Pi^{PV}_{ph}(q)(g'_{NN} + g'_{\Delta \Delta} - 2g'_{N \Delta}) 
\Pi_{\Delta h}(q)}
{(q^2 + g'_{NN} \Pi^{PV}_{ph}(q))(q^2 + g'_{\Delta \Delta} 
\Pi_{\Delta h}(q)) - g'^2_{N \Delta}
\Pi^{PV}_{ph}(q) \cdot \Pi_{\Delta h}(q)}.
\end{eqnarray}
The magnitude of $g'_{NN}$, $g'_{N \Delta}$, $g'_{\Delta \Delta}$, however, 
is not well determined. Usually $g'_{NN}$ is larger than $g'_{N \Delta}$ and 
$g'_{\Delta \Delta}$$^{2, 10, 11}$). In nonrelativistic approach, $g'_{NN}$ =0.9
, $g'_{N \Delta}$ = $g'_{\Delta \Delta}$ = 0.6 in ref. $^2$), $g'_{NN}$ = 0.9
, $g'_{N \Delta}$ = 0.38, $g'_{\Delta \Delta}$ = 0.35 in ref. $^{10}$). But in 
the relativistic approach,  $g'_{NN}$ = $g'_{N \Delta}$ = $g'_{\Delta \Delta}$ 
=0.6 is used in the calculation ref. $^{11}$), we adopt this prescription in this
paper.  

\section{Results and discussion}
In this section, we will give the numerical results for the real and the 
imaginary part
of the polarization insertion and pion dispersion relation. The nucleon
effective mass in the formula is determined by the Walecka's model in the
relativistic Hartree approximation ({\it RHA}), namely,  
${\tilde m}_{N}\over {m_{N}}$ = 0.72 for nuclear matter density $\rho = 
\rho_0$ ($k_F$ = 1.42 fm$^{-1}$). A remark on the effective mass of 
$\Delta$-isobars
$\tilde m_{\Delta}$ is as follows: The origin of the effective
mass of $\Delta$-isobars comes from the interaction between $\sigma$ meson
and $\Delta$-isobars$^{11, 16}$) (This statement is model dependent, there are 
other sources probably.). 
Due to this interaction, the $\Delta$-isobars will contribute to
the binding energy as well as the effective mass of nucleon at saturation 
density in the 1 - loop level or higher orders. 
 Since the nucleon effective mass is determined in the {\it RHA}
without $\Delta$-isobars, so for consistency we choose $\tilde m_{\Delta} =
m_{\Delta}$ in the numerical calculation. 
There will be no any other free parameters.

Shown in Fig. 1 is the real part of the polarization insertion function. They 
\begin{center}
\begin{tabular}{|c|} \hline
Figure 1 \\ \hline
\end{tabular}
\end{center}
are all negative in this case. In the limit ${\bf q} \rightarrow 0$, 
$\Pi^{PV}_{ph} \rightarrow 0$ is as expected, but $\Pi_{\Delta h}$ does not 
vanish. This nonvanishing value will contribute to the pion dispersion 
relation at the point ${\bf q} \rightarrow 0$. Another evident feature we can 
observe is that the {\it ph} excitation contribution is suppressed when short 
range correlation effect sets in. This suppression can also been seen 
in the imaginary part of the polarization insertion.

The imaginary part of -$\Pi'$ at density $\rho_0$ ($k_F$ = 1.42 fm$^{-1}$)
is shown in Fig. 2 and Fig. 3. In Fig. 2, -$\Pi'$ is shown as a function of
\begin{center}
\begin{tabular}{|c|c|} \hline
Figure 2 &Figure 3\\ \hline
\end{tabular}
\end{center}
energy $\omega$ for fixed momentum $q = 2.5 k_F$ (approximately 5$m_\pi$).
The first peak which starts from $\omega \sim 0.4 m_\pi$ ends at $\omega
\sim 3.2 m_\pi$ corresponds to the {\it ph} excitation, the $\Delta h$ 
excitation is represented by the second peak starts at $\omega \sim 4 m_\pi$
ends at $\omega \sim 6 m_\pi$, these values correspond to the boundary of 
the hatch region at $q \sim 5 m_\pi$ in Fig. 4. It is evident that the short 
range correlation effect suppresses the {\it ph} excitation but enhances the
$\Delta h$ excitation. This is the same conclusion we have drawn from Fig. 1.
The same is true for Fig. 3 where $\omega = 5 m_\pi$ is fixed, and the first 
peak (dotted and solid line curves) stands for $\Delta h$ excitation while the 
second peak (dotted and solid line curves) stands for {\it ph} excitation.
There is no zero value between peaks in this case, because the {\it ph} and
$\Delta h$ excitation overlap for $\omega = 5 m_\pi$, $q$ in 7$m_\pi \sim 
8 m_\pi$ as shown in Fig. 4. Again, the $\Delta h$ excitation is enhanced
and the {\it ph} excitation is suppressed by the short range correlation.

The pion dispersion relation in nuclear matter is determined by the trajectory 
of the poles of pion propagator, that is:
\begin{eqnarray}
\omega^2 - q^{2} - m^{2}_{\pi} - \mbox {Re} \Pi'(\omega, q) = 0.
\end{eqnarray}
here Re means taking the real part of $\Pi' (\omega, q)$, it includes
the contributions both from the real and the imaginary part of $\Pi^{PV}_{ph}$
and $\Pi_{\Delta h}$.  Firstly, we calculate the dispersion relation in case 
without short range correlation effect, that is $g'$ = 0, the result is shown
in Fig. 4. We can see that there are three disconnected curves indicated by 
\begin{center}
\begin{tabular}{|c|} \hline
Figure 4 \\ \hline
\end{tabular}
\end{center}
letters {\bf a, b, c}. The existence of curve {\bf a} implies the pion 
condensation happens at relatively larger momentum at about 5.6$m_\pi$. The
upper part of curve {\bf b} is pion-like pion mode, the lower part starts
from about $q \sim 0.3 m_\pi$ which is mainly due to {\it ph} excitation, is 
noncollective {\it ph}$^{10}$). These two modes connect at higher momentum and
energy. The noncollective {\it ph} excitation mode falls into the hatched 
region completely thus it is damped. The curve {\bf c} is noncollective
$\Delta h$ excitation mode, because it is mainly due to $\Delta h$ excitation
contribution. It disappears at $q \sim 2.6 m_\pi$.

When short range correlation sets in, the dispersion relation changes very much
as shown in Fig. 5. There is no counterpart of curve {\bf e} and the lower 
part of curve {\bf f}
\begin{center}
\begin{tabular}{|c|} \hline
Figure 5 \\ \hline
\end{tabular}
\end{center}
in Fig. 4, and new modes appear also. The curve {\bf f} is the compressed
one of curve {\bf c} in Fig. 4 by short range correlation. The short range 
correlation also brings about the new modes curve {\bf d} and the upper part
of curve {\bf g} which are damped by $\Delta h$ decays. 

We have seen the short range correlation effect is very larger indeed,
however, it is not well defined on the other hand$^{12}$). An alternative way
to construct the short range correlation to the polarization insertion is
as follows$^{8, 12}$:
\begin{eqnarray}
\Pi'(q) = {\bf q}^{4} \frac{\Pi^{PV}_{ph}(q) + \Pi_{\Delta h}(q)}
{(- {\bf q}^{2} + g' \Pi^{PV}_{ph}(q))(- {\bf q}^{2} + 
g' \Pi_{\Delta h}(q)) - g'^{2} \Pi^{PV}_{ph}(q) \cdot \Pi_{\Delta h}(q)},
\end{eqnarray}
that is to replace the $q^2$ in eq. (19) by -$q^2$. Substitute eq. (21) into
eq. (20), we get the dispersion relation shown in Fig. 6. In this case, the
\begin{center}
\begin{tabular}{|c|} \hline
Figure 6 \\ \hline
\end{tabular}
\end{center}
pion mode starts at $\omega = 1 m_\pi$, because the total polarization 
vanishes when {\bf q} $\rightarrow 0$. The free pion dispersion relation
(dashed line curve) above the full pion mode implies that the pion effective
mass in nuclear matter is reduced comparing to the free one as a function of
energy and momentum. The small island in the figure is mainly due to 
noncollective $\Delta h$ excitation and short range correlation. Compare
Fig. 5 with Fig. 6, we can see that the short range correlation has a very
large effect on the pion dispersion relation. Our studies also show that the
pion dispersion relation is very sensitive to other parameters in the model,
such as the effective mass of nucleon and $\Delta$-isobars, Landau-Migdal 
parameters and nuclear density.

The spectral function of pion $A_\pi$ in a $\Delta h$ model at zero and finite 
temperature in the nonrelativistic approach was calculated approximately$^8$)
and self-consistantly$^9$). It is defined as follows:
\begin{eqnarray}
A_\pi(\omega, q) = - {1\over\pi} {\mbox Im}\frac{1}{\omega^2 - q^2 -
m_\pi^2 - \Pi'(\omega, q)},
\end{eqnarray}
here the polarization insertion including short range correlation effect is 
given by eq. (19) and eq. (20). $A_\pi(\omega, q)$ plotted as functions
of $\omega$ for fixed $q = 2.5k_F$
is shown in Fig. 7. Similar to Fig. 2, for polarization eq. (19), the short
\begin{center}
\begin{tabular}{|c|} \hline
Figure 7 \\ \hline
\end{tabular}
\end{center}
range correlation suppresses the {\it ph} but enhance the $\Delta h$ 
excitation. The zero points in the denominator of eq. (22) gives the peaks
in spectral function. But for the polarization eq. (20), the peaks disappear
for those energy and momentum.
For {\it ph} excitation, its effect is similar to eq. (19), but for $\Delta
h$ excitation the short range correlation effect is very small. So it's very 
clear that the effect of short range correlation depends  on its
forms, and this effect can change the pion dispersion relation and other
observable very much.

\section{Summary and conclusions}
In summary of this study, we have derived the real $ph$ and $\Delta h$ 
excitation polarization
insertion for pion propagator by employing particle, hole, antiparticle
propagators representation of nucleon and $\Delta$-isobars, instead of their
"Density dependent" and "Feynman" propagators. 
The $ph$ and $N \bar N$, $\Delta h$ and $\Delta \bar N$, $\bar \Delta N$ 
excitations are separated. The real and the imaginary part of the total 
polarization insertion including {\it ph}, $\Delta h$ excitations and short
range correlation is calculated. We found that the short range correlation
largely suppresses the {\it ph} excitation but enhances the $\Delta h$
excitation very much. 
The pion dispersion relations are quite different from those calculated 
in the nonrelativistic approximation$^{2,8}$. The short range correlation
has a large effect on all branches of the pion dispersion relation.
The  pion condensation 
at momentum  $q \sim 5.6 m_\pi$ disappears when short range correlation
sets in. 

We have also studied the effect of the different forms of short range
correlation on the pion dispersion relation and pion spectral function. The
difference between two forms of short range correlation is very large. 
We have found but have not shown that the pion dispersion relation depends
very much on the effective mass of nucleon and $\Delta$-isobar and Landau-Migdal
parameters. 
These problem suggests that the short range correlation itself needs a careful
study, and only until it is well defined, we can have the definite answer
to the pion dispersion relation.

The author Liu is very grateful to E. Oset, Huan-Qing Chiang and Cheng-Guang 
Bao for very valuable discussions. 

\newpage
\section*{Reference}
\begin{description}
\item
1) C. Gale {\scriptsize AND} J. Kopusta, Phys. Rev. $\bf C35$(1987) 2107
\item
2) L. H. Xia, C. M. Ko, L. Xiong {\scriptsize AND} J. Q. Wu, 
   Nucl. Phys. $\bf A485$(1988) 721
\item
3)G. Chanfray {\scriptsize AND} P. Schuck, Nucl. Phys. {\bf A555}(1993) 329
\item
4) C. Y. Wong, {\it Introduction to High - Energy Heavy - ion Collisions},
  (World Scientific, Singapore, 1994)
\item
5) E. Oset, H. Toki {\scriptsize AND} W. Weise, Phys. Rept. 
   ${\bf 83}$(1982) 281
\item
6) C. Garcia - Pecio, E. Oset, L. L. Salcedo, Phys. Rev. 
   ${\bf C 37}$(1988) 194
\item
7) A. B. Migdal, E. E. Saperstein, M. A. Troitsky {\scriptsize AND} 
   D. N. Voskresensky, Phys. Rep. $\bf 192$(1990) 179
\item
8) P. A. Henning, H. Umezawa, Nucl. Phys. {\bf A571}(1994) 617
\item
9) C. L. Korpa, R. Malfliet, Phys. Rev. {\bf C52}(1995) 2756
\item
10)J. Helgesson {\scriptsize AND} J. Randrup, Phys. Rev. C52(1995)427
\item
11) T. Herbert, K. Wehrberger, {\scriptsize AND} F. Beck, Nucl. Phys. 
   ${\bf A 541}$ (1993) 699
\item
12) L. G. Liu, Phys. Rev. {\bf C51}(1995)3421
\item
13) L. G. Liu, Q. X. Ma, {\scriptsize AND} H. Q. Chiang,  High Energy Phys. 
   \& Nucl. Phys. {\bf 18}(1994)391
\item
14) L. G. Liu, Q. F. Zhou {\scriptsize AND} T. S. Lai, Phys. Rev. {\bf C51}
   (1995)R2302
\item         
15) L. G. Liu, {\it The binding energy of relativistic particle-hole
in nuclear matter in the relativistic $\sigma - \omega$ model}, 
Jou. Phys. (London) {\bf G22} (1996) 1799
\item
16) L. G. Liu, Q. F. Zhou, {\it The study of the binding energy of nuclear
matter in the relativistic $\sigma - \omega - \pi$ model with 
$\Delta$-isobar degree of freedom}, to be appear in Z. Phys. {\bf A}
\item
17) B. D. Serot {\scriptsize AND} J. D. Walecka, Adv. in Nucl. Phys. 
   {\bf 16}(1986) 1
\item
18) W. Bentz, A. Arima, H. Hyuga, K. Shimizu, {\scriptsize AND} K. Yazaki,
    Nucl. Phys. {\bf A 436}(1985) 593
\item
19) K. Wehrberger, Phys. Rept. {\bf 225}(1993)272
\item
20) see, e. g., H. B. Tang {\scriptsize AND} P. J. Ellis, {\it Redundancy
of $\Delta$-isobar parameters in effective field theories}, preprint
NUC-MINN-96/4-T, and references therein.
\item
21) L. M. Nath, B. Etemadi {\scriptsize AND} J. P. Kimel, 
    Phys. Rev. {\bf D3}(1971) 2153
\end{description}

\newpage
\section*{Figure Captions}
\begin{description}
\item
[Figure 1.]
The real part of the polarization insertion $- Re \Pi (\omega = 0, q)$
(in unit of $m_\pi^2$) as a function of momentum $q$ (in unit of $m_\pi$)
for density $\rho_0$. The {\it ph} and $\Delta h$ contributions are dotted
and dashed line. "ph + $\Delta$h + SRC" is the total result by eq. (19).
\item
[Figure 2.]
The imaginary part of - $\Pi' (\omega, q)$ (in unit of $m^2_\pi$) as
a function  of $\omega$ (in unit of $m_\pi$) for q=2.5$k_F$. $g'$ = 
0, dotted line; $g'$ = 0.6, solid line. 
\item
[Figure. 3.]
- $\Pi' (\omega = 5m_\pi, q)$ (in unit of $m^2_\pi$) versus $q$ 
(in unit of $m_\pi$) at density
$\rho_0$ and $g'$ = 0 (dotted line), and $g'$ = 0.6 (solid line).
\item
[Figure 4.]
Pion dispersion relation in  nuclear matter in the case $\rho = \rho_0, g'
$ = 0. The axes are in unit of $m_\pi$. The hatched regions indicate the region
where the imaginary part does not vanish: the lower part is for {\it ph}
excitation the upper part is for $\Delta h$ excitation. 
\item
[Figure 5.]
The same as Fig. 4 but $g' = 0.6$. The letters indicate the corresponding
dotted lines.
\item
[Figure 6.]
Pion dispersion relation by eq. (20, 21), $g' = 0.6$. 
The dashed line is for free pion.
\item
[Figure 7.]
Pion spectral function $A_\pi(\omega, q)$ (in the unit of $m_\pi^2$) versus
$\omega$ (in unit of $m_\pi$), for $q = 2.5 k_F$, $\rho = \rho_0$.
Dotted line $g'$ = 0; solid line $g'$ = 0.6, the polarization by eq. (19); 
dashed line $g'$ = 0.6 and the polarization by eq. (21).
\end{description}
\end{document}